# High Current Proton Tests of the Fermilab Linac

M. Popovic, L. Allen, A. Moretti, E. McCrory, C.W. Schmidt and T. Sullivan

Fermi National Accelerator Laboratory[1]

Batavia, Illinois, USA

*Abstract*

The peak current limit for the Fermilab Linac was recently studied. The purpose was to learn what components of the present Linac can be used for the first stage of a proposed proton driver[1]. For this application the Linac must provide a $H^-$ beam in excess of 5000 mA-μsec per pulse. The original Fermilab Linac was designed for protons with a peak current of 75 mA and a pulse length of four Booster turns (~10 μsec). The high energy replacement was designed for a peak current of 35 mA and a beam pulse length of 50 μsec. The present $H^-$ source cannot deliver more than ~80 mA which produces 55 mA in the Linac. Using a proton source allows the system to be tested to currents of ~100 mA and pulse lengths long enough to observe the effects of long pulses. This test has shown that the present Linac can accelerate beam having a peak current up to ~85 mA with beam loss comparable to the present Linac operation (~45 mA). The results of the test will be presented.

## 1 INTRODUCTION

During its lifetime the Fermilab Linac has gone through two mayor modifications. In both cases these improvements were motivated by the need for higher beam intensity from the Booster synchrotron. Construction of the Linac began in 1968 and a 200-MeV proton beam was first produced on November 30, 1970. The design goal[2] of 75 mA and 10 μsec (four Booster turns) was achieved quickly and surpassed. Although the design intensity was 75 mA, the Linac was built to accelerate at least 100 mA with similar beam emittance. Emittance preservation is essential for successful horizontal injection and stacking of four turns in the Booster. Eight years later the proton source was replaced by a $H^-$ source to accelerate a long, low-intensity $H^-$ beam of 25 mA and build intensity in the Booster using multi-turn charge-exchange injection. In the summer[3] of 1993 the Linac was upgraded again. The last four drift-tube tanks were removed and a side-coupled structure installed to increase the final energy to 400 MeV. The higher injection energy in the Booster increased the magnetic guide field at injection, reduced the frequency range of the RF accelerating system and increased the Booster's space-charge limit at injection and therefore its possible intensity. Although the Linac energy upgrade was designed for a peak beam current of 35 mA, typical current at the end of the Linac was between 28 and 36 mA. Since the energy upgrade there have been small changes in operating parameters of the ion source, low energy transport line and Linac that have resulted in a steady increase in peak beam current extracted from the Linac. Figure 1 shows the peak beam current in the Linac and corresponding Booster beam intensity over the past thirty years of operations. In the constant quest for higher beam intensity and considering that the side-coupled structure was design for a maximum beam current of 35 mA, the intensity at 400 MeV has been greatly increased.

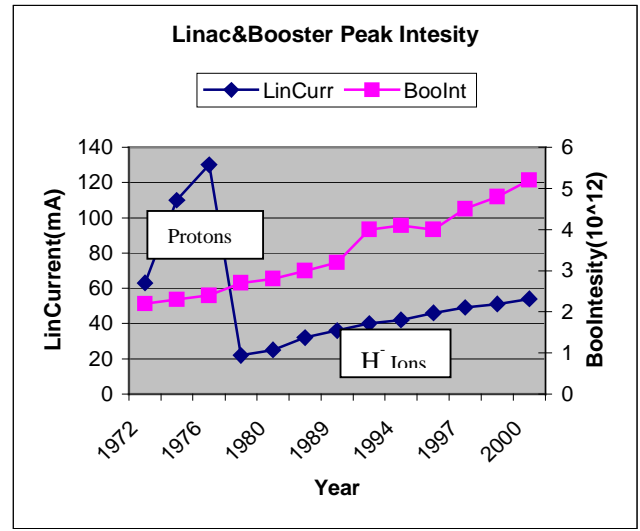

**Figure 1. Linac and Booster intensity with time**

## 2 SYSTEM MODIFICATION

Obtaining the anticipated current using an $H^-$ source would have required a significant source development program which, in part, was the purpose of this study. Therefore an old proton source, a duoplasmatron, was reinstalled in one of the Cockcroft-Walton pre-accelerators. This source once produced several hundred milliamperes for short pulse (3-5 μsec) 200-MeV Linac operation and could easily provide a proton beam with a current for this test. Other modifications required changing the polarity of the preaccelerator high voltage and magnetic transport dipoles. Also the 750-keV input and 400-MeV output transport lines had to be retuned for protons. The 750-keV transfer line is short. It has only three quad triplets and a Buncher cavity. The polarity of the quads were kept as for $H^-$ operation. A Trace2D

---

[1] This work is supported by U.S. Dept. Of Energy through the University Research Association under contract DE-AC35-89ER40486

model was used to tune the line, see Figure 2. This model was also used to show that the line has sufficient flexibility to match to the Linac with no change in the Linac polarities or settings.

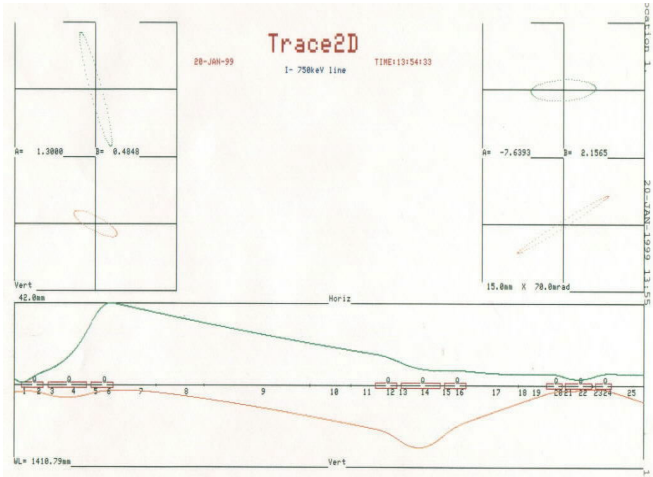

Figure 2. Trace2D run of the 750-keV transport line for protons

## 3 COURSE OF TEST

H⁻ operating conditions for beam transmission and capture in the low-energy linac is 74% and 95% in the high-energy linac. Thus there is a loss of ~30% that must be accommodated by the source.

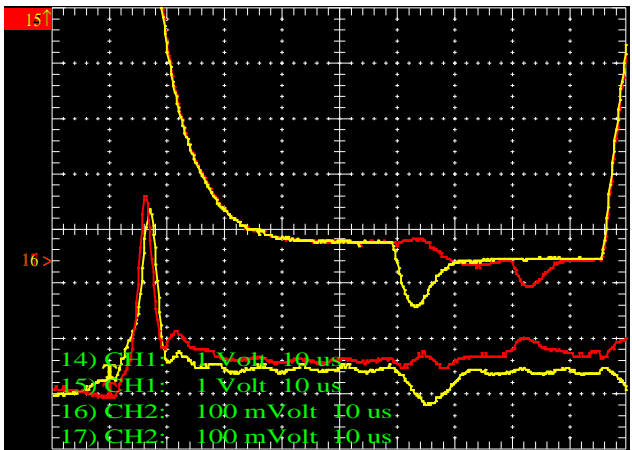

Figure 3. Loading of the High-Energy Klystron RF pulse.

It was assumed that all limits on beam current will be visible for a beam pulse between 10 to 30 μsec. Sparking rates in the side-couple modules were monitored for any increase, as were the amplitudes and reflected power on the each RF station. The beam loss monitors were carefully watched at all times. All high peak current related measurements were done with a minimal number of pulses. Figure 3 shows the RF gradient envelope and reflected power signals for Station 7 of the side-coupled linac. Red traces are RF signals with beam. Yellow traces are the same signals without beam. The bumps are a result of gain and feed-forward adjustments. The feed-forward correction compensates for beam loading starting at the head of the beam pulse and lasting through the duration of the pulse.

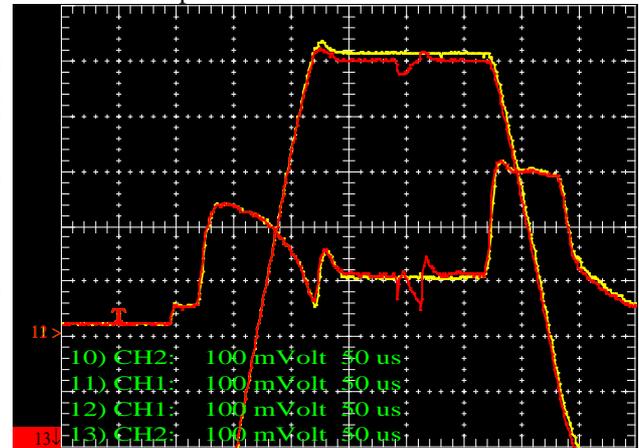

Figure 4. Loading of Low-energy RF pulse.

These signal were watched for signs of RF saturation during high peak current running. Similar RF signals for the low-energy linac tanks were watched. Figure 4 shows a low-energy gradient and reflected power signal with and without beam. This portion of the Linac was originally design for higher peak currents. The dip in the signal is present only during beam time and is a result of beam loading and a lack of full beam loading compensation. Every attempt is made to keep the gradient signal constant during beam time.

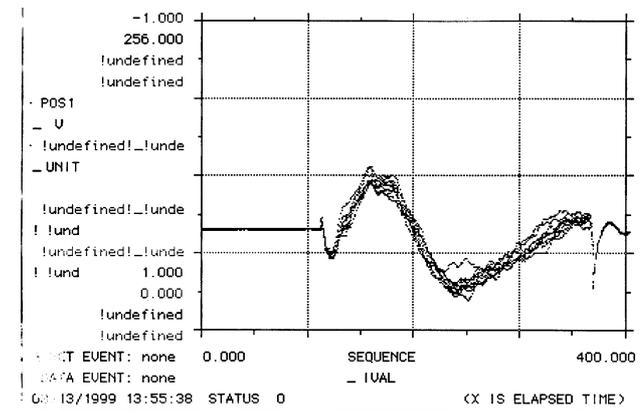

Figure 5. Horizontal beam motion at 400 MeV.

Energy variation during the pulse can be observed at the end of the Linac following a spectrometer magnet. These variations are believed due to the variation of the gradients along the drift-tube linac. Figure 5 shows the horizontal position of the beam during the pulse. To insure that the emittance of the high current beam from the source was not degraded, the beam emittance at the entrance to the Linac was measured, see Table 1. Clearly, the beam emittance is rather constant and not significantly dependant on the beam current for values between 78 and 92 mA. This is important because there was significant change in the beam loss going from 85 to 95 mA or higher. Figure 6 shows toroids and beam loss monitors along the Linac for a current of ~85 mA.

| LinacInputEmittanceForProtonStudy | | | |
|---|---|---|---|
| 3/18/99 | E.McCrory | | |
| | EmitPrT1In | LinacOut(mA) | E(95%Nor) |
| | Horizontal | 78 | 2.4 |
| | | 82 | 2.4 |
| Emittance in pi-mm-mr | | 88 | 2.8 |
| | | 92 | 2.4 |
| | BuncherOFF | | 2.2 |
| | Vertical | 78 | 3.1 |
| | | 82 | 2.9 |
| | | 88 | 2.9 |
| | | 92 | 3 |
| | BuncherOFF | | 3.1 |

Table 1. High current beam emittance.

Figure 7 shows the same signals for a current of ~95 mA. The loss along the Linac has increased. Wire profiles along the high-energy linac have not shown any visible increase in beam size for the transverse planes.

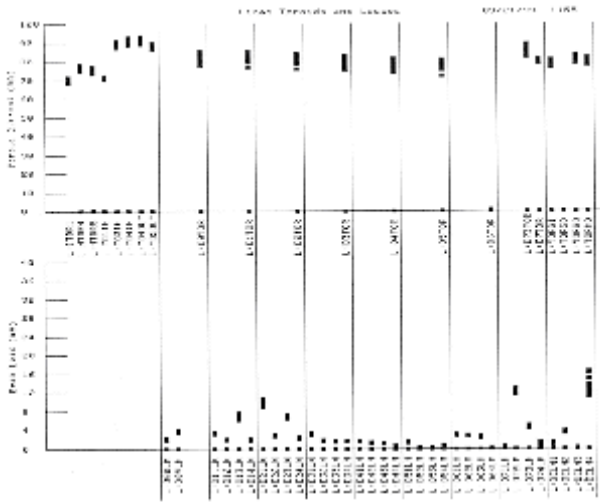

Figure 6. Current and losses through Linac at ~85 mA.

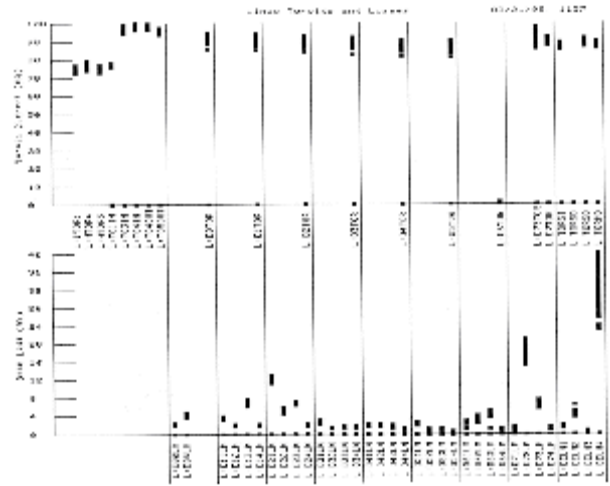

Figure 7. Current and losses through Linac at ~95 mA.

## 3 SUMMARY

A proton beam with a peak current up to 90 mA was accelerated through the Linac with similar losses as a lower intensity $H^-$ beam. For currents above 90 mA there is additional loss with indications that this loss is related to a large energy spread and lack of RF voltage to properly accelerate the beam and keep it in the bucket. It is believe that the present Fermilab Linac can accelerate up to 90 mA of $H^-$ beam for future uses with a suitable source.

## 4 ACKNOWLEDGEMENTS

The authors would like to acknowledge the work of several people who were instrumental in carrying out this test. Although the test was relatively simple the preparation and restoration of the Linac was extensive. From the Linac, James Wendt and Ray Hren prepared the source, its installation and later removal; Lester Wahl assisted with RF monitoring and control. From the Mechanical group, Danny Douglas, Mike Ziomek and Ben Ogert prepared and restored the Cockcroft-Walton.